\documentstyle[aps,preprint]{revtex}

\begin{document}
\begin{center}
{\large{\bf Relativistic mean-field study of  light nuclei near drip line}}\\

B. K. Agrawal$^1$, Tapas Sil$^1$, S. K. Samaddar$^1$ and J. N. De$^{2}$\\
$^1$Saha Institute of Nuclear Physics, 1/AF Bidhannagar, Calcutta 700 064, India\\
$^2$Variable Energy Cyclotron Centre, 1/AF Bidhannagar, Calcutta 700 064, India
\end{center}

\begin{abstract}
The relativistic mean field theory is applied to study some exotic
properties of neutron rich nuclei as recently observed,  namely, 
extension of  the drip-line for $F$ nuclei from $^{29}F$ to $^{31}F$ and
the appearence of a new shell closure at neutron number $N=16$. We find $^{31}F$
to be bound against one-neutron dripping but unbound only marginally for
two neutron separation. The calculated functional 
dependence of one-neutron separation
energy with neutron number  for different values of $T_Z = (N-Z)/2$ signals
a new shell closure at $N=16$ for neutron rich nuclei with $T_Z\ge 3$. This
is further corroborated from the study of the deformation and the gap
across the Fermi surface in these light nuclei.
\end{abstract}
\vspace*{0.5 true in}
PACS Number(s): 21.60.-n, 21.30.Fe,27.30+t\\
Keywords: Relativistic mean-field, drip line, shell closure
\newpage
\section{Introduction}

Recent  experiments with radioactive ion beams have mapped
out neutron drip line nuclei around $N=20$ and $N=28$ \cite{sak1,sak2}.
These exotic nuclei with high isospins have many peculiar properties quite
distinct from those found in the stability valley. The usual shell closures
at $N=20$ and 28 have been questioned, new shell closures have been
predicted and  doubts have been expressed for the validity in this 
region of the 
usual nucleon-nucleon interaction or nuclear matrix elements which are 
so successful in describing the nuclei close to the
stability line. The neutron-rich nuclei near the drip line show large
neutron skin or neutron halo and the radii are larger than those
expected from the usual $r_0A^{1/3}$ law. A number of "Borromean" nuclear
systems like $^{11}Li$, $^{31}F$ , $^{32}Ne$ etc. have been discovered which
result in unstable nuclei after one neutron removal but stable ones on
two neutron removal. These are only a few of the exotic properties 
found in nuclei near the drip line. 

There have been a number of attempts, both in the non-relativistic as well
as in the relativistic frameworks, to understand the properties of
nuclei in the vicinity of the drip line. In different models, 
though some of the properties are
qualitatively the same, there are also some striking differences. As for
example, experimental results \cite{mot} show that $^{32}Mg$ is highly
deformed ($\beta_2 \sim 0.5$) whereas the RMF theory predicts \cite{lal1}
nearly spherical configuration for this isotope; the shell model
calculations with relatively smaller model space also have similar
conclusions \cite{cau1}. On the other hand, a recent shell model
calculation carried out in the extended model space \cite{cau2} produce
deformation in consonance with the experimental finding.
However, some anomalies persist  in the binding energies of some nuclei 
around the drip line. 

Several calculations of drip line nuclei around $N=28$ have been performed
applying the RMF theory in the Hartree-BCS \cite{ren} as well as within the
Hartree-Bogoliubov \cite{lal2} framework.
The coordinate space description of the RMF theory in the spherical
configuration has been used \cite{lal3} for the study of neutron rich
nuclei around $N=20$.  With the advancement of radioactive ion beam 
techniques, the neutron drip line is
getting shifted to higher $N-$value. As for example, a recent experiment
at RIKEN \cite{sak2} shows that $^{31}F$ rather than $^{29}F$ is the most
likely candidate for the drip line fluorine isotope. It is conjectured
\cite{sak2} that onset of deformation is responsible for the extrastability
for $^{31}F$.

Recent analyses \cite{oza} of the single-neutron separation
energies $S_n$ and  other  experimental data for neutron-rich light nuclei,
however, indicate the appearance of a new magic number at N=16 and the
disappearance of the traditional magicity at N=20. Dissolution of the N=8
magic number in neutron-rich $Li$ and $Be$ isotopes \cite{sim,kel} and the
appearance of magicity at N=6 in $^{8}He$ \cite{ots} have also been pointed
out. The structures of these nuclei near the drip line are discussed in
reference to the picture of the island of inversion \cite{war} and have
recently been systematically studied in the framework of Monte-Carlo shell
model (MCSM) \cite{uts,cau2}. In the present work, we will focus on the
theoretical understanding of the extension of the drip line in $F$ chain and
the appearance of shell closure at N=16 in the light of relativistic
mean field framework.

The organization of the paper is as follows. In Sec. II, the formalism used
have been out-lined very briefly. In Sec. III, the results and discussions
are given. Finally, in Sec. IV, we present the summary and the concluding
remarks.

\section{Formalism}

The details of the formalism of the RMF theory for calculating various
observables like deformation, total energy etc. can be found in Refs.
\cite{gam,agr}. However, for the sake of completeness we write down the
effective Lagrangian density used describing 
the nucleon-meson many-body systems
followed by a brief discussion. The  Lagrangian density reads
\begin{eqnarray}
\label{lag}
{\cal L}&=& \bar\Psi_i\left ( i\gamma^\mu \partial_\mu - M\right )\Psi_i
+ \frac{1}{2} \partial^\mu\sigma\partial_\mu\sigma - U(\sigma)
- g_\sigma \bar\Psi_i \sigma\Psi_i\nonumber\\
&& - \frac{1}{4}\Omega^{\mu\nu}\Omega_{\mu\nu}
+\frac{1}{2}m_\omega^2\omega^\mu \omega_\mu - g_\omega \bar\Psi_i \gamma^\mu
\omega_\mu\Psi_i 
-\frac{1}{4}\vec{R}^{\mu\nu} \vec{R}_{\mu\nu} + \frac{1}{2} m_\rho^2 \vec{\rho}^\mu\vec{\rho}_\mu\nonumber\\
&&- g_\rho \bar\Psi_i \gamma^\mu\vec{\rho}_\mu\vec{\tau}\Psi_i
-\frac{1}{4}F^{\mu\nu}F_{\mu\nu} - e\bar\Psi_i \gamma^\mu \frac{(1-\tau_3)}{2} A_\mu\Psi_i
.
\label{Lag}
\end{eqnarray}

The scalar-isoscalar meson $\sigma$, the vector-isoscalar meson $\omega$
and vector-isovector meson $\rho$ are included in this description. The
arrows in Eq. (1) indicate isovector quantities. The scalar
self-interaction term $U(\sigma)$ of the $\sigma$ meson is taken to be
non-linear,
\begin{equation}
U(\sigma) = \frac{1}{2}m_\sigma^2 \sigma^2 + \frac{1}{3}g_2\sigma^3
+\frac{1}{4}g_3\sigma^4
.
\end{equation}
The nucleon mass is $M$; meson masses are $m_\sigma$, $m_\omega$ and
$m_\rho$; $g_\sigma$, $g_\omega$ and $g_\rho$ are the corresponding
coupling constants and $e^2/4\pi = 1/137$ is the fine structure constant.
The field tensors for $\omega$ and $\rho$ are given by $\Omega^{\mu\nu}$
and $\vec{R}^{\mu\nu}$; for the electromagnetic field it is $F^{\mu\nu}$.
Recourse to the variational principle followed by the mean-field
approximation treating the fields as $c-$numbers results in  the Dirac
equation for the nucleon and Klein-Gordon type equations for the mesons and
the photon. For the static case, along with the time reversal invariance
and charge conservation, the equations get simplified. The resulting
equations known as the relativistic Hartree equation or RMF equation along
with  BCS approximation for inclusion of pairing are solved
self-consistenetly to yield the fields and the single-particle states in an
axially deformed basis expansion method \cite{rin}. 

The solution of the fields in the axially deformed basis for the odd-even
or odd-odd nuclei is quite a difficult task. The time reversal symmetry
in the mean field is violated for these systems. 
The odd particle induces polarisation current and time-odd component in the
meson fields. The time-odd component plays important role in the description
of magnetic moment and moment of inertia in rotating nuclei \cite{hof,afa}.
However, the effect on deformation and binding energies are very small and
can be neglected to a good approximation \cite{lal4}.
In the present calculation
for odd nuclei, we employ the blocking approximation \cite{pat} which
restores the time reversal symmetry. 

\section{Results and Discussions}

For our calculation, 
we choose the  NL3 parameter set \cite{lal5} for the
Lagrangian density given by Eq. (1). This particular set reproduces the
bulk properties of nuclei quite satisfactorily including the nuclear matter
incompressibility $K_\infty$. To check the sensitivity of the results  with
the choice of the parameter set we have also calculated the ground state
properties of some of the isotopes near the drip line using 
NL1 and NLSH sets having significantly different $K_\infty$. Among these
three sets, the  results with the
NL3 set are in better agreement with the experimental values. In the
following we shall present the results calculated with the
NL3 parameter set. The
calculations have been performed by expanding fermionic and bosonic wave
functions in 20 oscillator shells for axially deformed configuration.
 The $p-p$ and
$n-n$ pairing correlations have been considered in the BCS approximation.
The strength of these correlations are determined from 
the proton and neutron pairing
gaps $\Delta_p$ and $\Delta_n$ evaluated from the experimental odd-even
mass differences in the 4-point formula.
The more accurate five-point formula \cite{ben} might have been
employed, however, near the drip lines since some points
would have been missed, for consistency we use the four-point
formula throughout; moreover, the difference between the two is
found to be very nominal. Though the experimental
masses for the relevant nuclei  are available \cite{aud} with large 
neutron excess, a few are still missing near the neutron drip line. 
The pairing
gaps for these nuclei are taken by extrapolating the general trends of
$\Delta_p$ and $\Delta_n$.

\subsection{Location of the $F$ and $Ne$ neutron drip }
We have performed calculations for the binding energies of 
$F$ and $Ne$ isotopes extending 
from the stability line to the drip line lying around $N=20$.
In Fig. 1 the proton 
(filled circles joined by dashed line) and neutron (open circles joined
by full line) pairing gaps for $Ne$ and $F$ are shown as a function of
neutron number as obtained from the 4-point formula. 
The circles are replaced by asterisks where the masses used are obtained 
from systematics \cite{aud}. The oscillations of
$\Delta$ with maxima at even $N$ and minima at odd $N$ are manifestations of
the weakening of pairing correlations for odd nucleon number\cite{boh}. It is
seen that for $F$, there is a tendency of saturation for $\Delta_p$ and
$\Delta_n$ at high values of $N$ which is not that prominent for $Ne$
isotopes. Beyond the experimentally available values, 
we take $\Delta_n= 1$ MeV for both
the systems except for $N=20$ where $\Delta_n=0$     
and for $\Delta_p$ we take 2 MeV and 3.5 MeV for $Ne$ and $F$,
respectively.  The probable uncertainties
for these choices may be $\sim$0.5 MeV as can be estimated from Fig. 1.

The calculated ground state binding energies for the different isotopes of $Ne$ and
$F$ along with the experimental binding energies are shown in Fig. 2. For
both the systems it is seen that the agreement of the calculated values
with the experimental ones are good upto $N=16$ beyond which deviation
starts building up. These deviations for $F$ isotopes are larger by about
a factor of 2  compared to those for $Ne$. It is difficult to gauge whether
such large deviations can be explained in the
relativistic Hartree-Bogoliubov (RHB) theory \cite{lal2}.
It may be mentioned that the
agreement of the RMF binding energies with the experimental ones is very
good even close to the neutron drip line for isotopes with atomic numbers
$Z \ge 12$ \cite{lal1}. This indicates that possibly the 
correlation effects beyond
mean field play more important role for nuclei having neutron drip line
around $N=20$. However, since the uncertainties in the binding energy
differences between two successive isotopes  are much less, it is still
meaningful  to explore the drip-line characteristics for these nuclei. 

In Fig. 3, we display the one-neutron separation energies $S_n$ for the
isotopes of $Ne$ and $F$ where the filled circles joined by the dashed line
correspond to  the experimental $S_n$ \cite{aud} and
the open circles joined by full line represent the 
calculated values with the NL3 parameter set.
It is seen that the odd-even oscillations are similar to the observed ones 
for the two
nuclei, however, there is some quantitative disagreement, particularly for
$F$ with $N > 15$. The calculated two neutron separation 
energies $S_{2n}$ for the
two isotope chains are shown in Fig. 4 along with the available
experimental values \cite{aud}. 
The stability of all the $Ne$ isotopes agrees with the
experimental findings, particularly, the stability of 
$^{29}Ne$ and the instability of $^{31}Ne$.
It may be pointed out that $^{29}Ne$ is found to be unstable in 
the shell-model calculation \cite{cau1} contrary to the experimental finding.
In our calculations, $^{29}F$ and $^{34}Ne$ lie on the drip line.
The nucleus $^{31}F$ is found to be
bound against one-neutron dripping but unbound by a few hundred KeV against
two-neutron dripping contrary to the recent experimental observation
\cite{sak2} indicating that $^{31}F$ is a stable isotope. The pairing gaps
were varied by 0.5 MeV but that does not change our conclusion. 
It may be pointed out that a  very recent Monte-Carlo shell model (MCSM)
calculation \cite{uts} also gets $^{31}F$ unbound by a few hundred KeV.

Recently, it has been pointed out \cite{sak2} that the large 
deformation ($\beta_2 \sim
0.3$) is responsible for the extrastability of $^{31}F$.
To investigate
the origin of the large deformation near the drip line, 
we display the evolution of the 
neutron and proton single particle energy levels in Figs. 5 and 6,
respectively for the  $Ne$ and $F$ nuclei around $N=20$. The dashed 
lines in the figures correspond to the Fermi energies. 
From Fig.5 at $N=20$, the gap between the 1$d_{3/2}$ and 1$f_{7/2}$
orbitals is $\sim 6$ MeV. With the increasing neutron number ($N>20$) some
of the $m-$components of these orbitals come very close  to 
each other and thereby the lower $m$-components of the 
1$f_{7/2}$ orbital may behave like an intruder state; this induces large 
prolate deformation in both the systems.
The proton orbitals do not
show such behavior, this produces lower deformation for protons
compared to that for neutrons near the drip line. The deformations for the
$F$ and $Ne$ isotopes are displayed in Fig. 7 where these characteristics
are evident.

The rms radii for protons and neutrons for the nuclei $Ne$ and $F$ as a
function of neutron number are shown in Fig. 8. We note that the proton
radius increases slowly with the neutron number, whereas the neutron radius
follows approximately an $N^{1/3}$ law. We find some undulations in 
the proton radii,
particularly for the relatively lower neutron number. This may be traced back
to the oscillations of the pairing gaps shown in Fig. 1. Higher the value
of $\Delta$, levels above the Fermi surface get higher occupancies producing
larger radii. This is reflected in the radii shown in Fig. 8.

To study the effects of the different parameter sets on the location of the
drip line nuclei, we repeated the calculations with the NL1 
and the NLSH parameter
sets. Though the absolute binding energies of different nuclei may vary by
a few MeV depending on the parameter set, the conclusions regarding the 
drip line nuclei considerd here remain unchanged.

\subsection{Shell closure at N=16}

A shell closure for nuclei gives rise to a break in the plot of one neutron
separation energy ($S_n$) versus neutron number at fixed $T_Z$ (i.e.,
neutron execss) \cite{boh}. Recent experimental data \cite{oza,kan}
available for the $S_n$ clearly signals appearance of shell closure at
N=6, 16 and 32 and disappearance of the conventional shell closure at N=8 and
20 for large values of $T_Z$. In this subsection, we mainly focus on
the appearance of shell closure at $N=16$.
The calculated one-neutron separation energies
$S_n$ for odd-N and odd-Z nuclei for
$T_Z$ lying between 0 to 4 are displayed as a function of N in the upper
panel of Fig. 9; its lower panel shows $S_n$ for odd N and even Z nuclei with
$1/2 \le T_Z\le 9/2$. The odd-even effects are eliminated with this choice.
The magic number appears as a decrease  of $S_n$ with N \cite{boh}. 
The traditional
magic number at N=20 is found to persist in our calculation for all $T_Z$
(up to $T_Z=9/2$) though experimentally there are indications of the dissolution
of the magicity \cite{oza} at N=20 for $T_Z \ge 4$. 
At N=16, a clear break in $S_n$ is 
seen for $T_Z\ge 3$ suggesting the appearance of a new shell closure at this
neutron number.  These correspond to the nuclei $^{23}N$, $^{24}O$,
$^{25}F$ and $^{26}Ne$. At $T_Z = 1$ corresponding to $^{30}Si$ a break in
$S_n$ is also observed; this does not correspond to a true 
shell closure as discussed later. 

In order to identify the shell closure more clearly, we also study the
evolution of the deformation $\mid \beta_2 \mid $ and the
energy gap $\Delta\epsilon$ across the
Fermi surface (the energy difference between the last occupied orbit and
the first unoccupied one) with neutron number for different nuclei. In the
upper panel of Fig. 10, the deformation of the isotopes of the odd-Z nuclei
$N$, $F$ and $Na$ are shown; the same is displayed in the lower panel for
the even-Z nuclei $Ne$, $Mg$ and $Si$. All the isotopes of $O$ considered
are found to be spherical or nearly spherical; their deformations are
not displayed here. For all the nuclei considered, the deformation vanishes
at N=20 except for $F$ having a minimum 
in $\mid \beta_2 \mid $ with a small value.
At N=16, the isotopes of $N$ and $Ne$ are found to be spherical 
reinforcing the existence of
shell closure. For $F$, there is a local minimum in deformation at
N=16 with a relatively small value of $\mid \beta_2 \mid \sim 0.14$. 
Though there is
an apparent break in $S_n$ at N=16 for $Si$, no local minimum in $\beta_2$
is found at this neutron number. The deformations of $Na$ and
$Mg$ at N=16 are relatively large.

The energy gap across the Fermi surface of a 
traditional closed-shell nucleus is relatively large
compared to those of the neighbouring nuclei. In Fig. 11, the evolution of
this gap $\Delta \epsilon$ for a few odd-Z (upper panel) and even-Z (lower
panel) nuclei is displayed as a function of the neutron number. At N=20,
all the nuclei that are stable show robust peaks. At N=16, the nuclei with
$T_Z \ge 3$ also exhibit peaked structures. These peaks are relatively less
prominent compared to those at N=20. From the examination of one neutron
separation energy, deformation and shell gap, we observe an
extra-stability  for the nuclei with N=16 and with $T_Z \ge 3$; thus, 
the appearance of a new shell-closure at this N-value becomes evident.
It is however weaker compared to the traditional one at the nearby neutron 
number N=20 for all $T_Z$ considered here.

We have not included the effect of zero-point vibrations in our
calculations, which in general influences the ground state energy. However,
the parameters of the effective Lagrangian have been determined through
fitting the experimental ground state energies of some nuclei; further
inclusion of zero-point oscillation may thus lead to some double-counting.
The intrinsic state generated in the mean-field approximation is a
superposition of states of different angular momenta; projection of the
ground state of good angular momentum from this mixed state has in general
an influence on both the ground state energy and the deformation. The
angular momentum projection from the RMF state is a nontrivial task, we
have not yet attempted it.

\section{Summary and Conclusions}

We have calculated some recently observed ground state properties
for the isotopic chains extending from the 
stability line to the neutron drip line for several light nuclei
in the relativistic mean-field  theory.  It is found that  though the binding
energies of the isotopes near the drip line (around $N=20$) are
overestimated by a few MeV, the binding energy differences of neighbouring
nuclei are fairly reproduced.
The present calculation predicts $^{29}F$ and $^{34}Ne$
as drip line nuclei. Experimentally, till date $^{31}F$ is
considered as the drip-line nucleus and for $Ne$ upto mass number 32 has been
observed. In our calculation, $^{31}F$ is bound against one neutron
separation but unbound against two-neutron separation by only a few hundred KeV. 
The stability of $^{29}Ne$ and instability of $^{31}Ne$ found in our
calculations are in agreement with the experimental observation.

The behaviour of the one neutron separation energy $S_n$, the deformation
$\beta_2$ and the gap across the Fermi-surface of the neutron rich nuclei
around N=16 clearly indicate the existence of a neutron shell closure at
N=16 for $T_z \ge 3$. The dissolusion of the neutron shell closure at N=20
for $T_z \ge 4$ suggested from experiments is  however not observed in the
present model.

The present calculation suffers from a number of limitations. The
treatment of pairing in the BCS approximation is rather uncertain,
particularly for the exotic nuclei; the self-consistent Hartree-
Bogoliubov approach for pairing is preferable. Correlations beyond
the mean field may also be important in the present context.
In a recent shell-model calculation \cite{ots}, the mechanism for
the appearance of shell-closure at N=16 and also the disappearance
at N=20 for neutron-rich nuclei is attributed to the shift of
the $1d_{3/2}$ level to higher energies closer to the $1f_{7/2}$
orbit. The microscopic origin of the reorganisation of the
levels  is claimed to lie in the spin-isospin part of the nucleon-nucleon
interaction. In the present calculation, the level reorganisation
is not that marked because of which the energy gap $\Delta\epsilon$
is not too large at N=16 and the shell-closure persists at N=20
even for very neutron-rich nuclei. The probable incompleteness
of the present $\sigma - \omega - \rho$ version of the RMF theory
for the description of very exotic systems is manifest here
and is a possible pointer
to the necessity of the inclusion of the pionic degrees of freedom 
in the RMF framework.

\newpage
\noindent {\bf Figure Captions}
\begin{itemize}
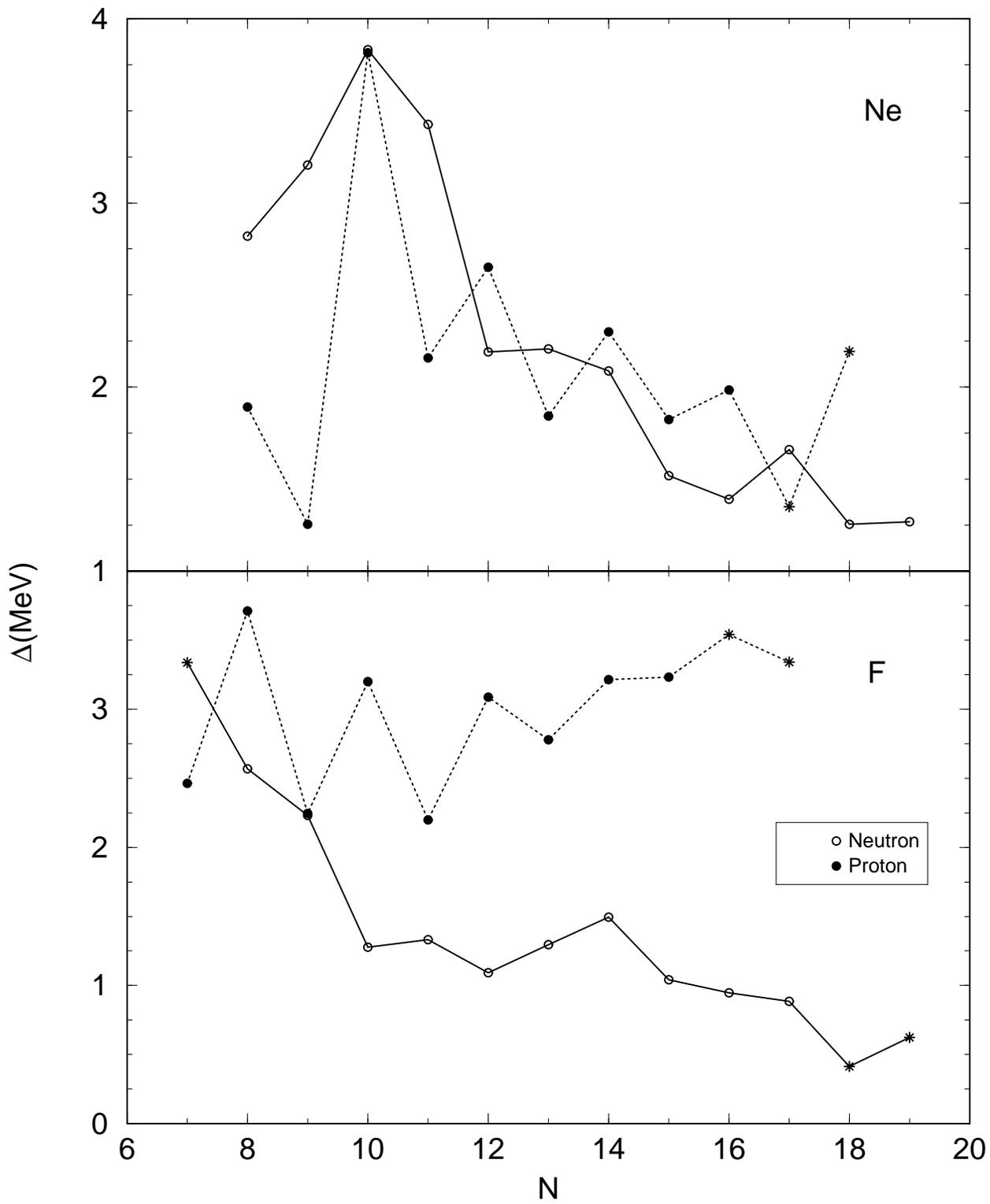
\item[Fig. 1] Experimental pairing gaps for neutron (open circles
joined by full line) and proton (filled circles joined by dashed line) 
for the isotopes of $Ne$ (top panel) and $F$ (bottom panel).
The asterisks stand for results obtained using masses from systematics.
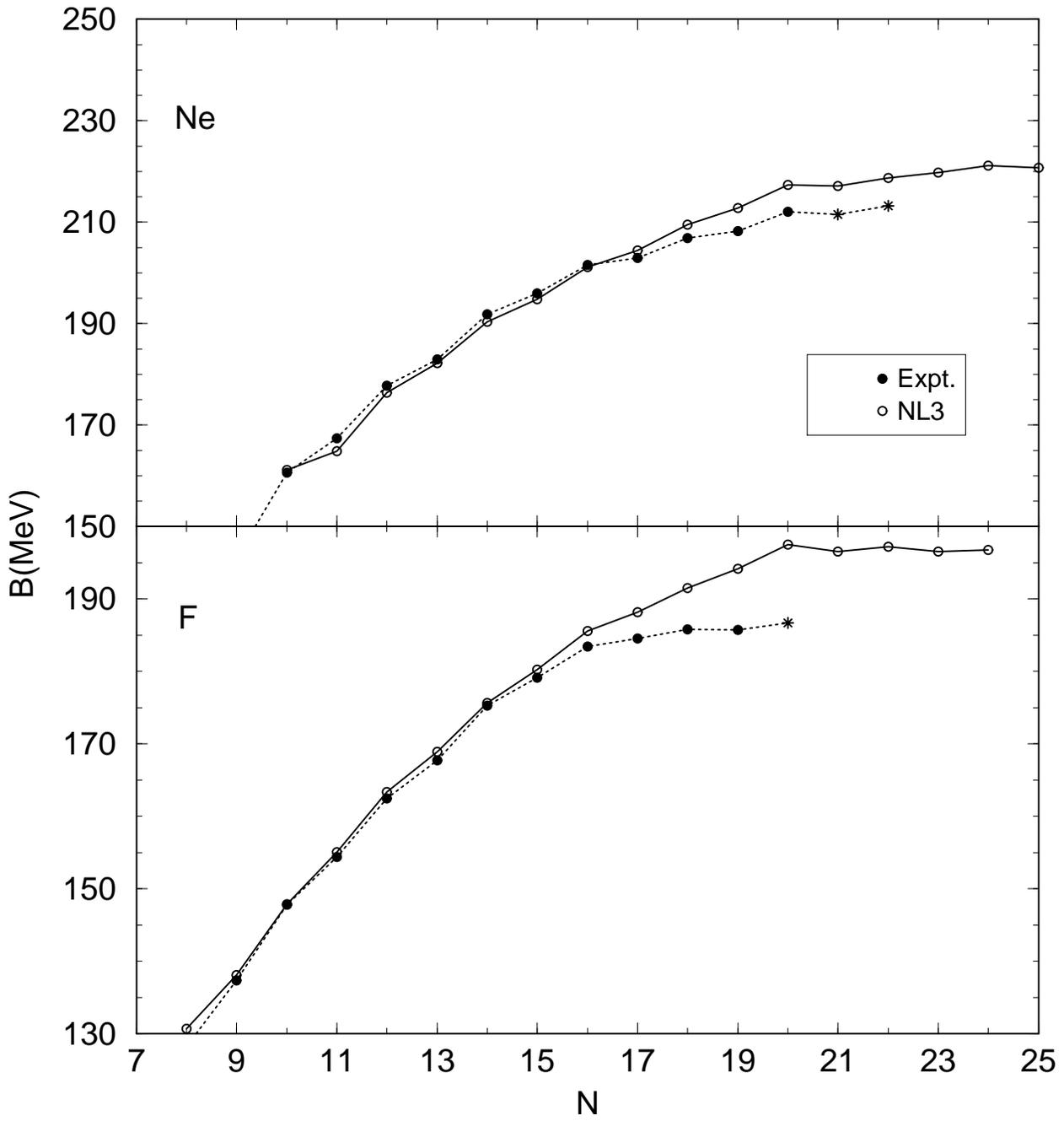
\item[Fig. 2] Calculated (with the NL3 parameter set) and experimental
binding energies for different isotopes of $Ne$ (top panel) and $F$ (bottom
panel). The asterisks stand for results obtained using masses from
systematics.
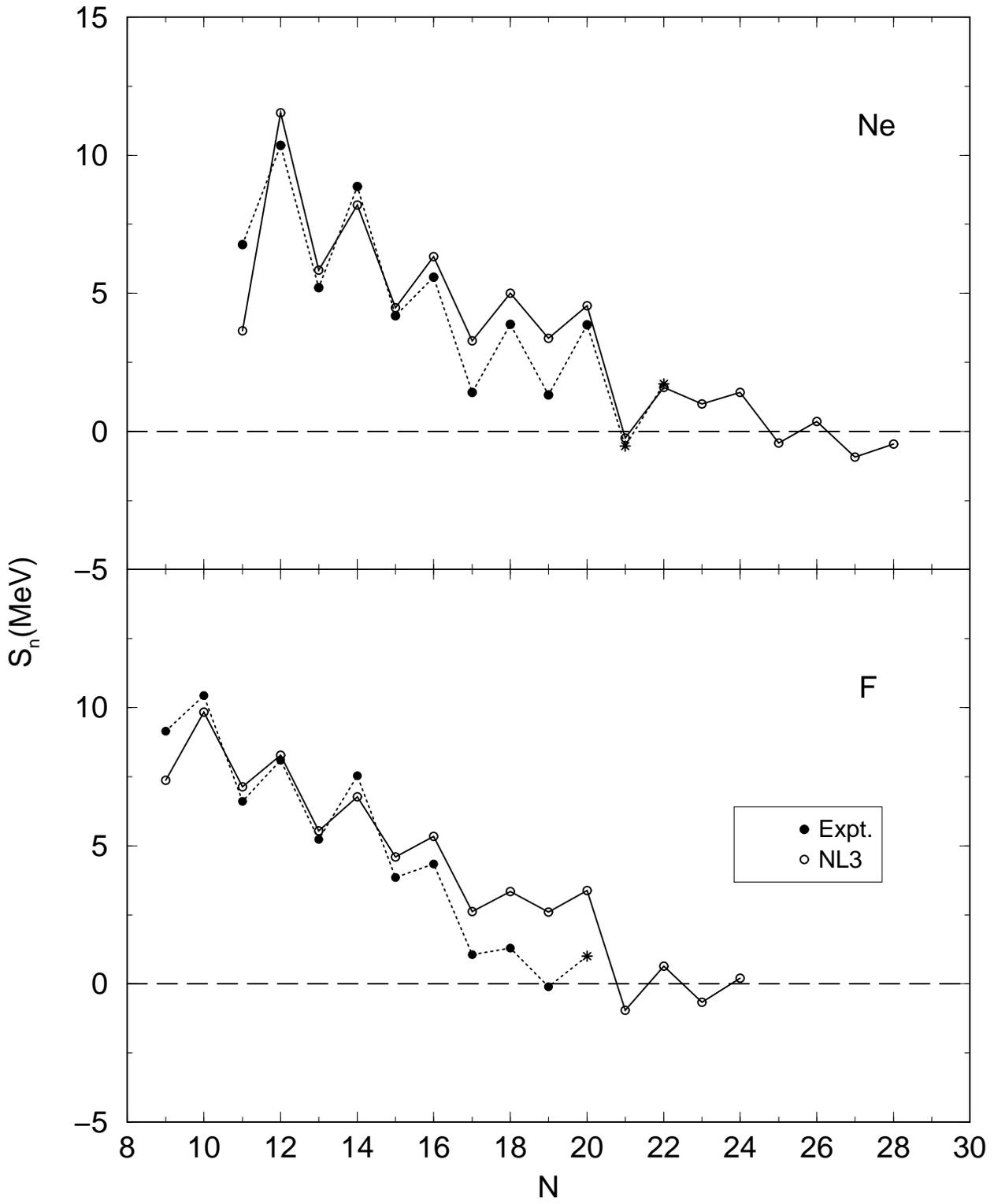
\item[Fig. 3] Calculated (open circles joined by full line) and experimental 
(filled circles joined by dashed line ) one-neutron separation 
energies $S_n$ for the different isotopes of
$Ne$ (top panel) and $F$ (bottom panel). Calculations are performed with
the NL3 parameter set. 
The asterisks stand for results obtained using masses from systematics.
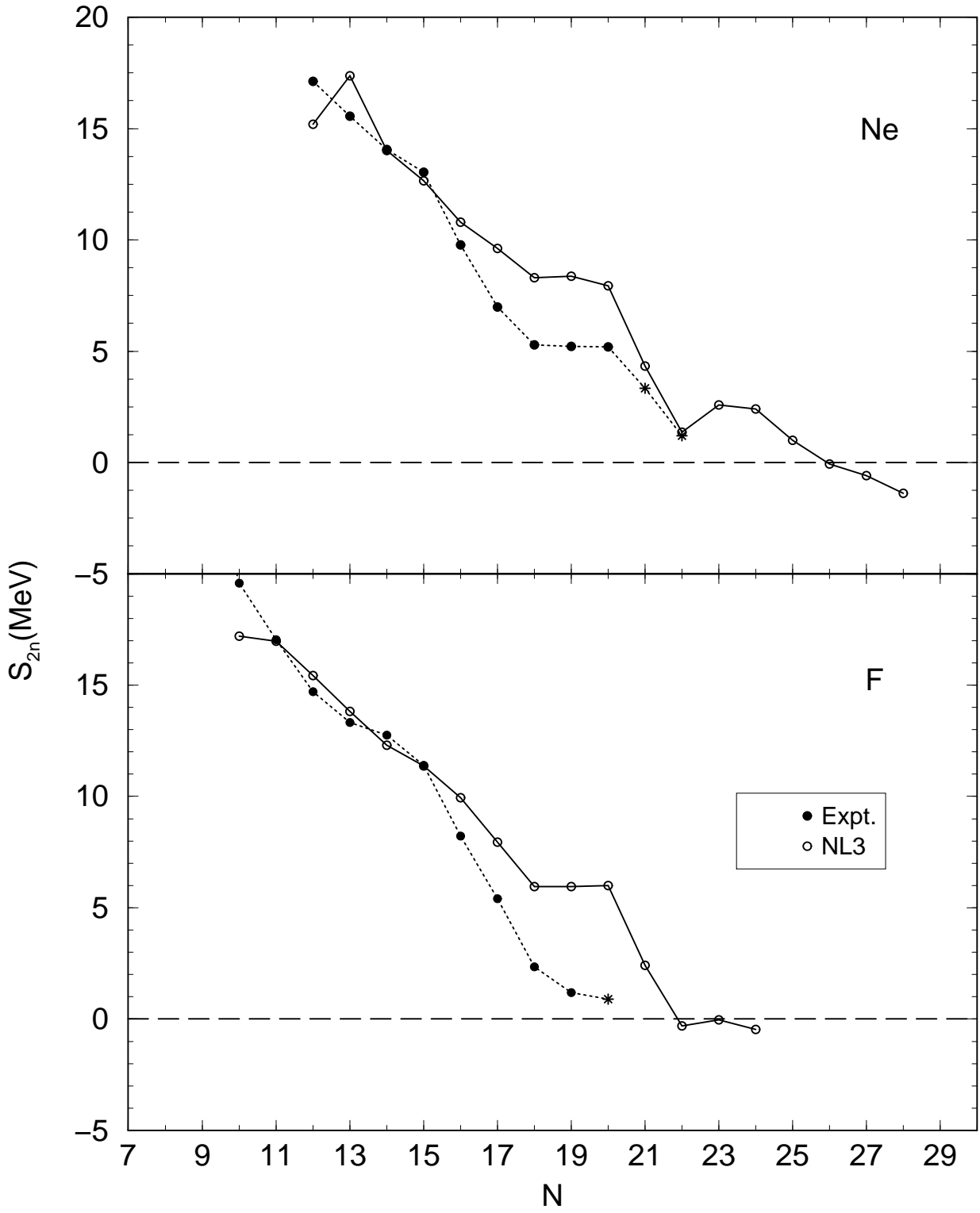
\item[Fig. 4] Same as in Fig. 3 for two-neutron separation energies $S_{2n}$.
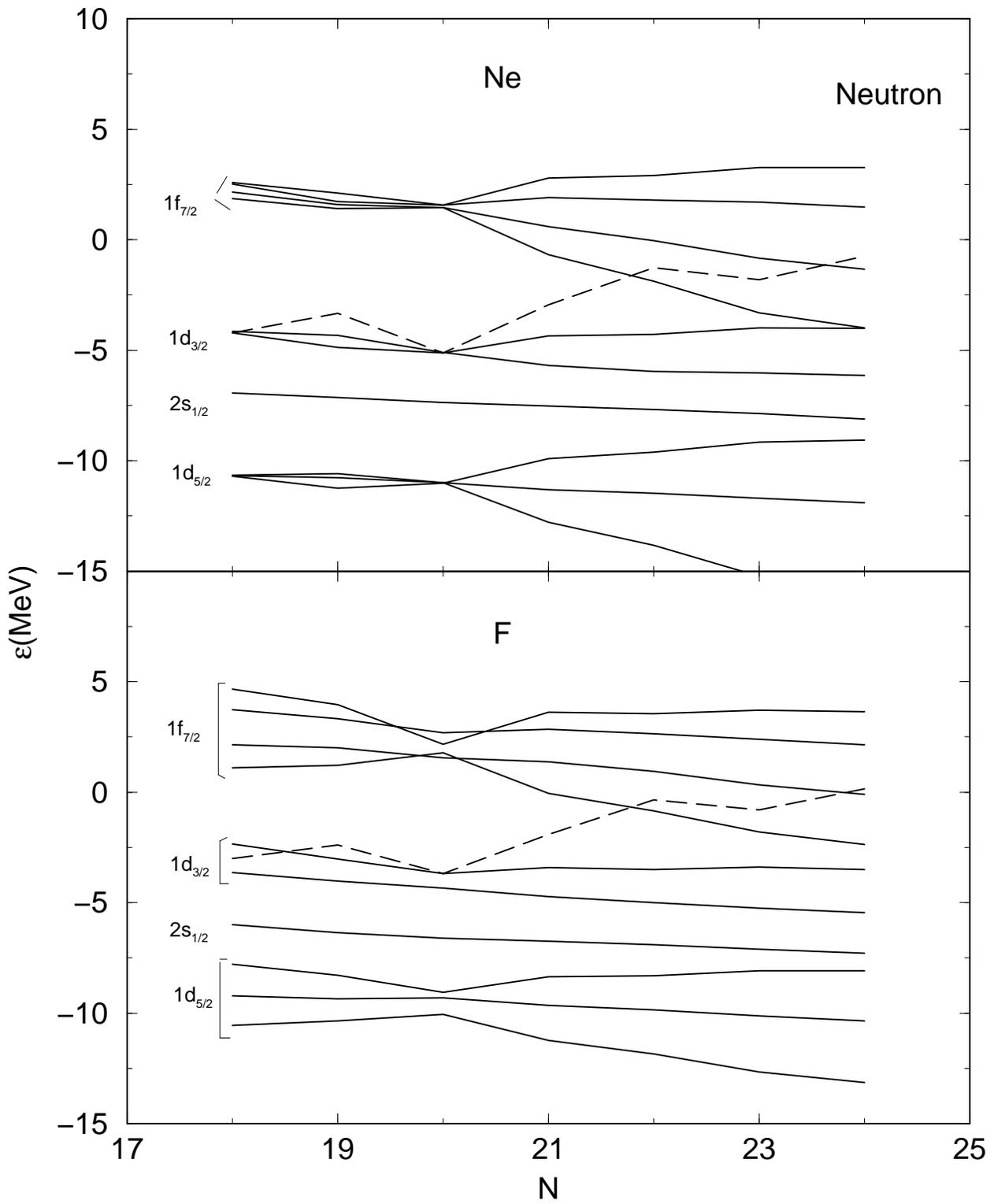
\item[Fig. 5] Neutron single-particle spectra around the Fermi surface
(dashed line) for different isotopes of $Ne$ (top panel) and $F$ (bottom
panel) close to the drip line calcualted with the NL3 parameter set. 
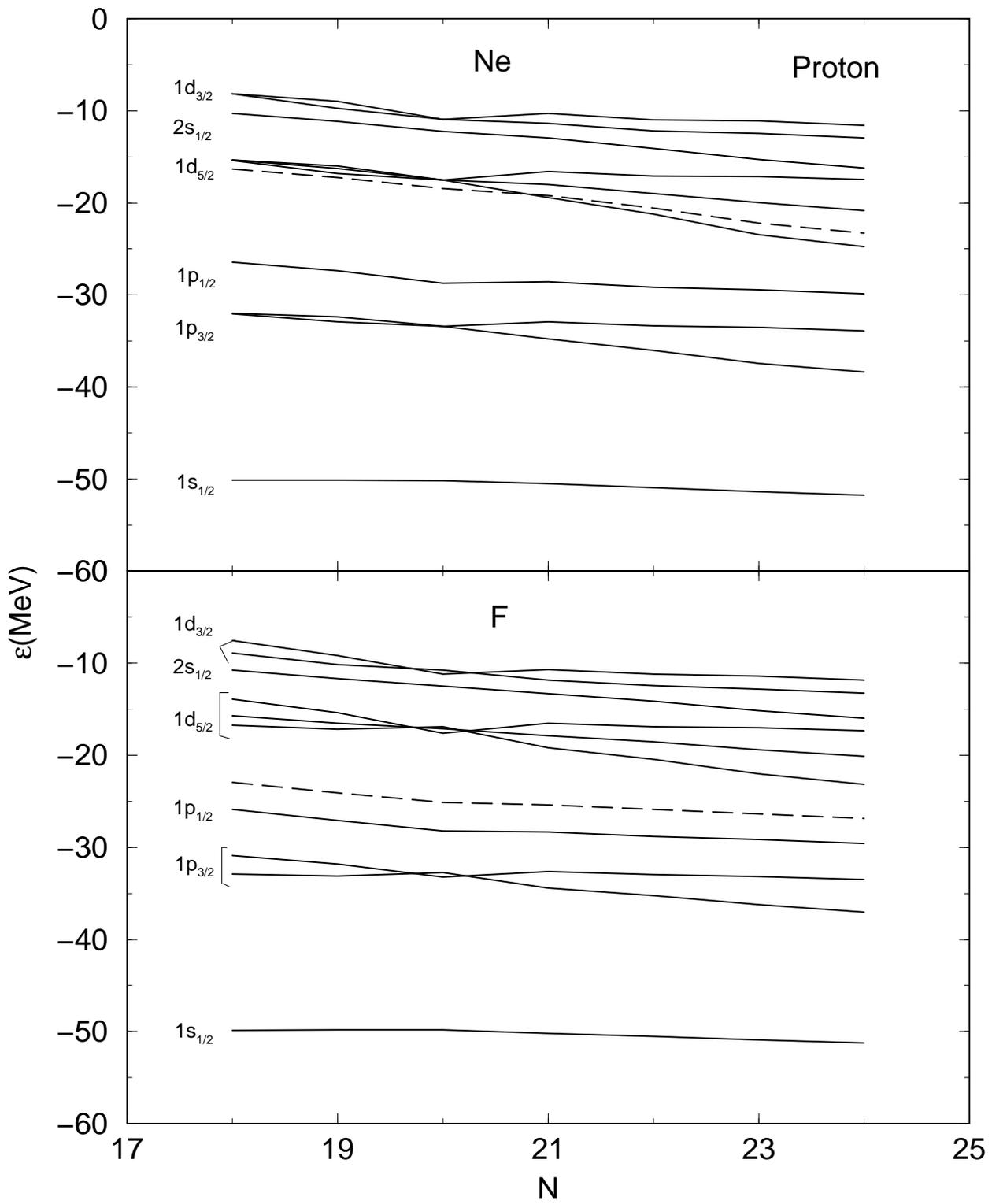
\item[Fig. 6] Same as in Fig. 6 for the proton spectra.
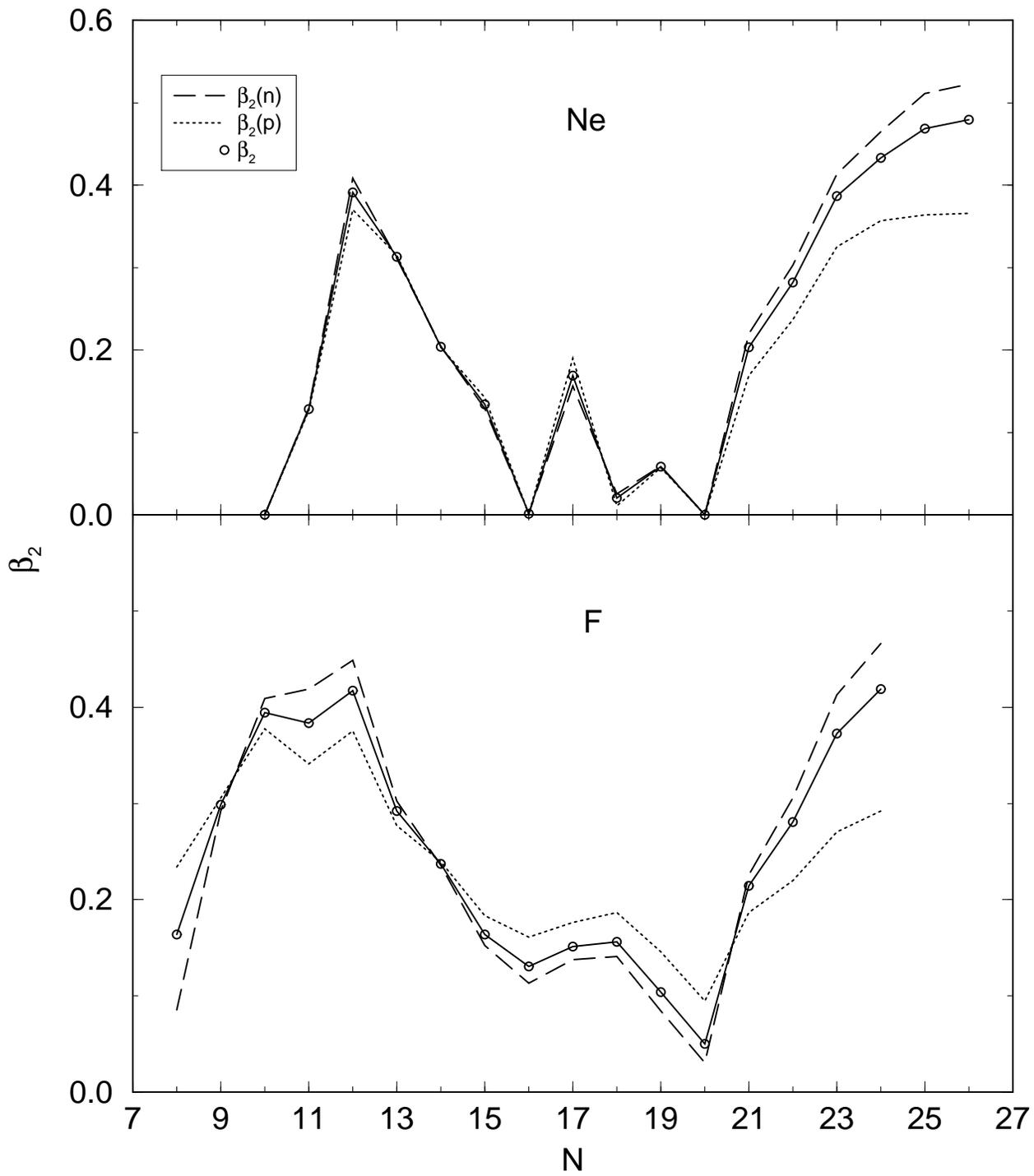
\item[Fig. 7] Neutron quadrupole deformation $\beta_2$(n) (dashed line) and
proton quadrupole deformation $\beta_2$(p) (dotted line) for the different
isotopes of $Ne$ (top panel) and $F$ (bottom panel)  calculated with the
NL3 parameter set. The open circles joined by 
full line represent the total deformation.
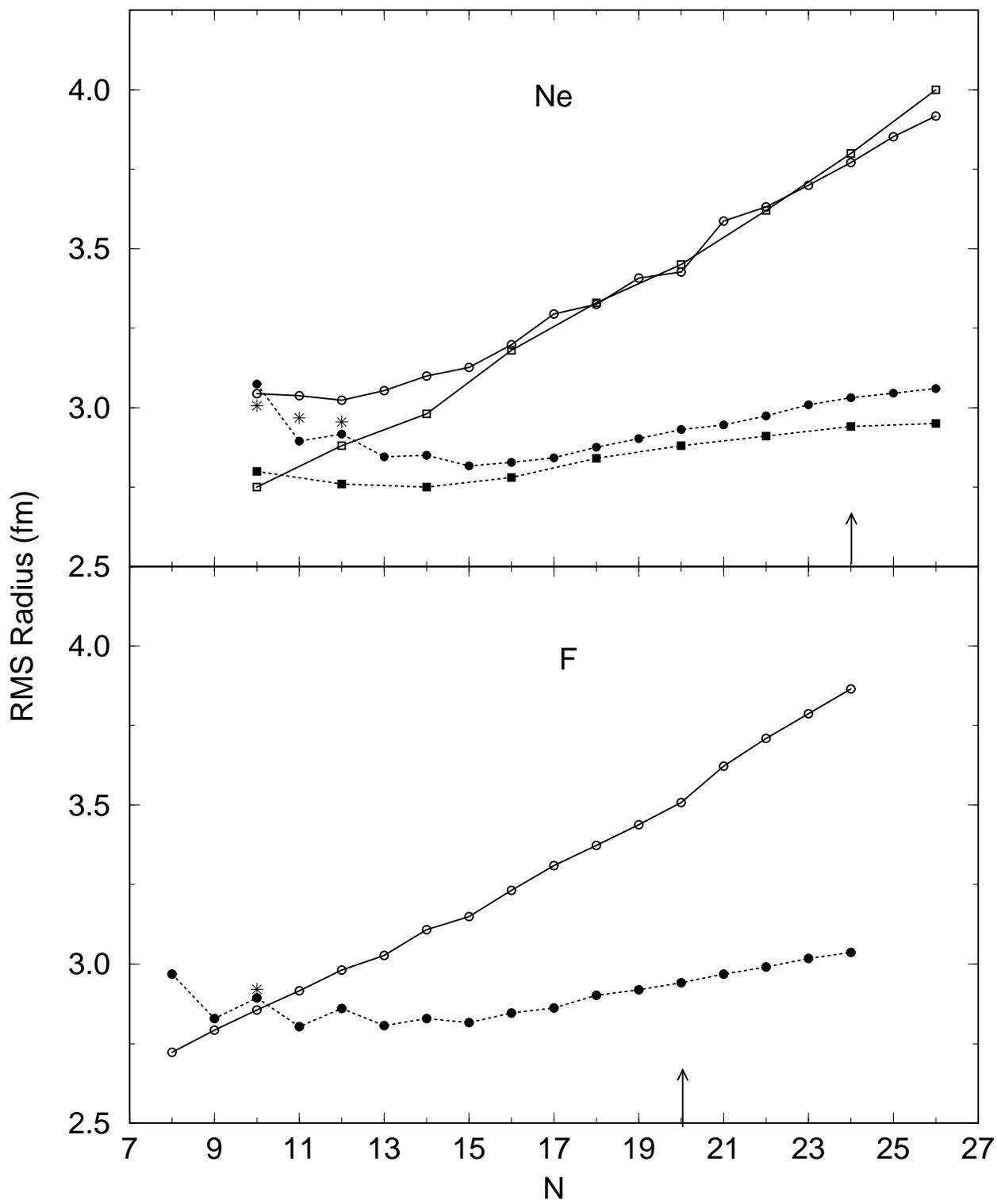
\item[Fig. 8] Neutron (open circles joined by full line) and 
proton (filled circles joined by dashed line) rms radii for
the different isotopes of $Ne$ (top panel) and $F$ (bottom panel)
calculated with the NL3 parameter set. The arrows on the abscissa
correspond to the locations of the neutron drip lines obtained
in the present calculation. The asterisk symbols refer to the
experimental proton rms radii. In the upper panel, the open and
filled squares joined by full and dashed lines, respectively, refer
to the neutron and proton rms radii from Ref.[9].
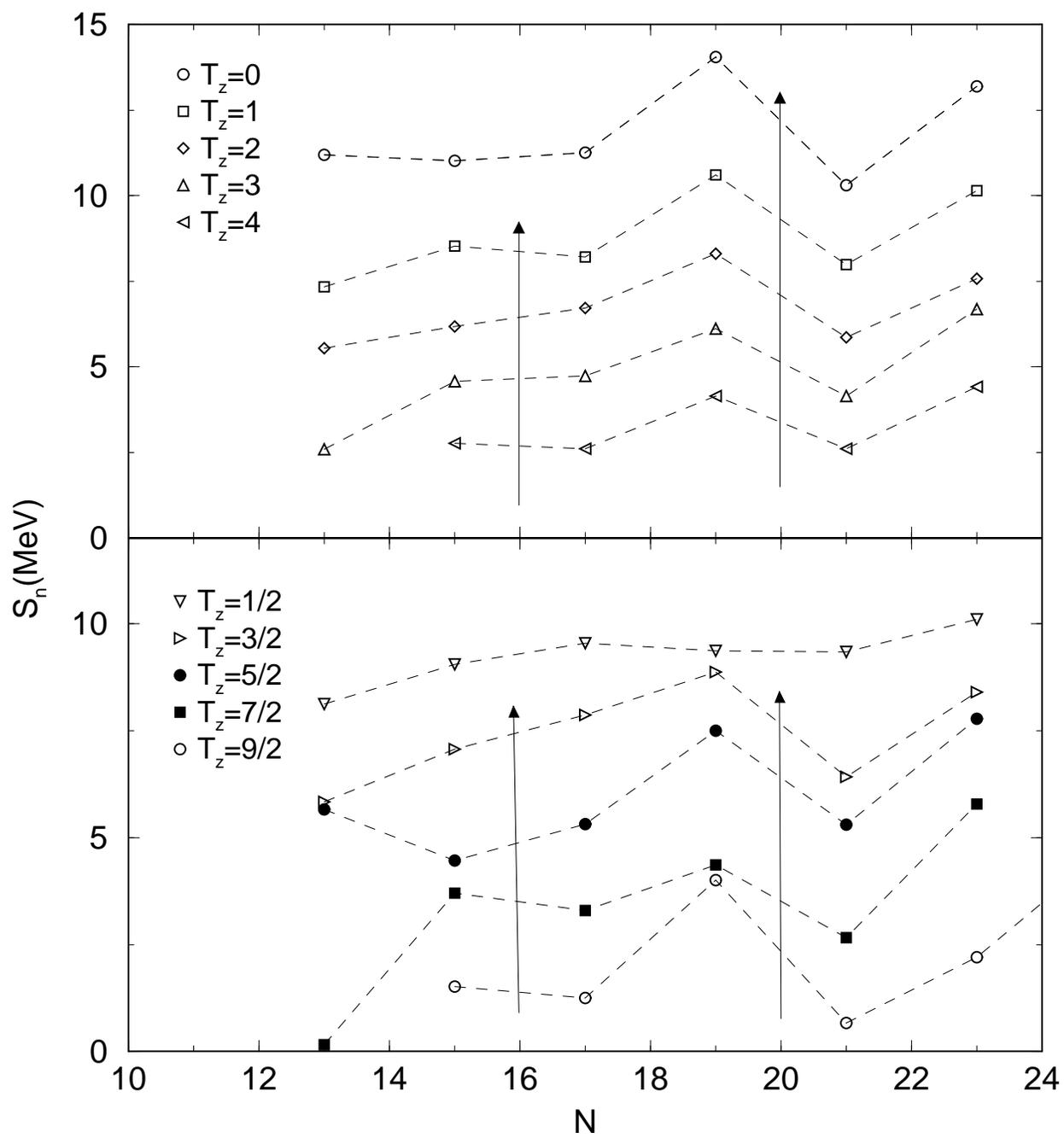
\item[Fig. 9] One neutron separation energy ($S_n$) as a function of
neutron number for odd-odd (upper panel) and odd-even (lower panel) nuclei.
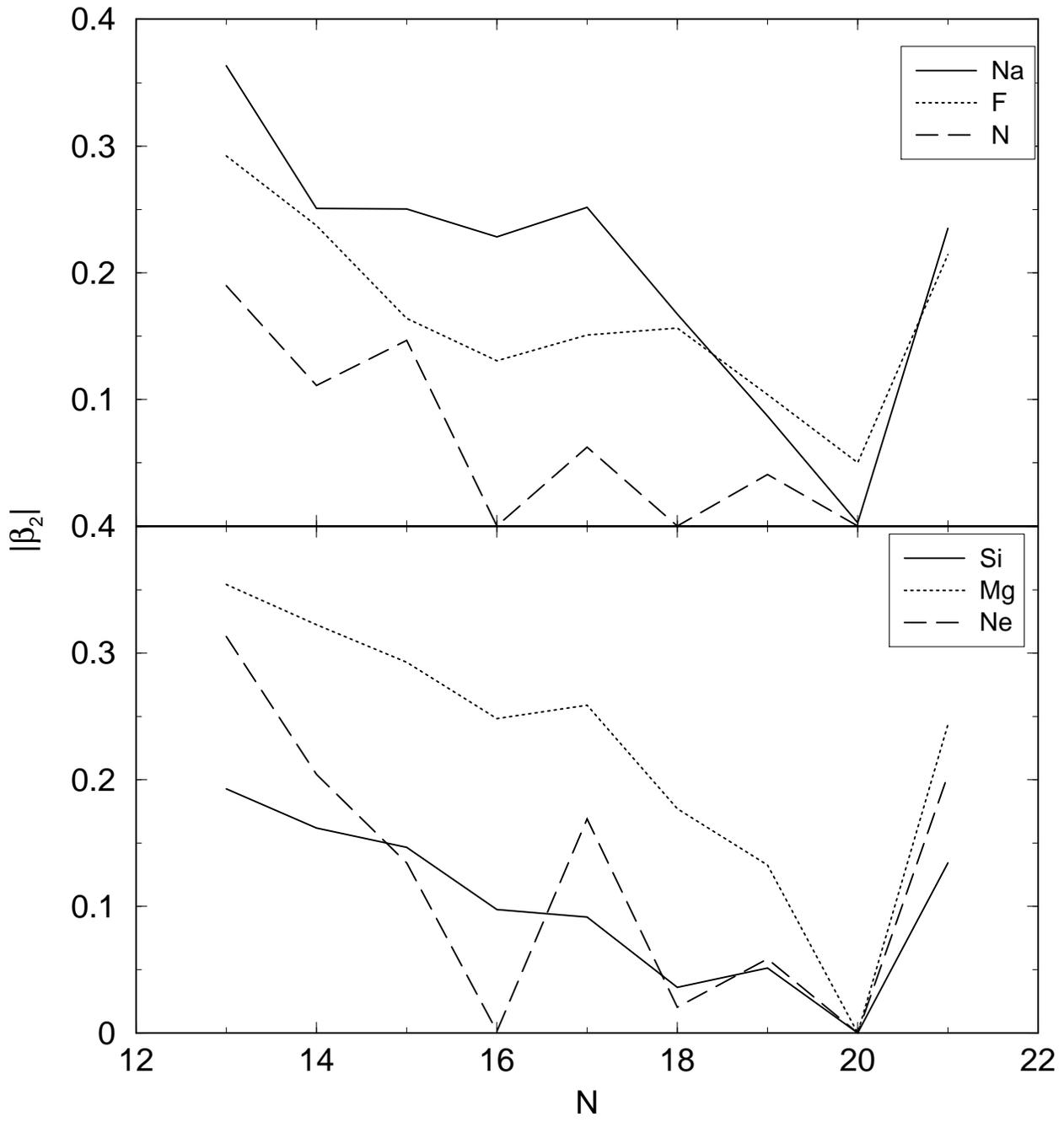
\item[Fig. 10] The magnitude of the ground state deformation $\mid\beta_2\mid$
as a function of neutron number for a few odd-Z (upper panel) and even-Z
(lower panel) nuclei.
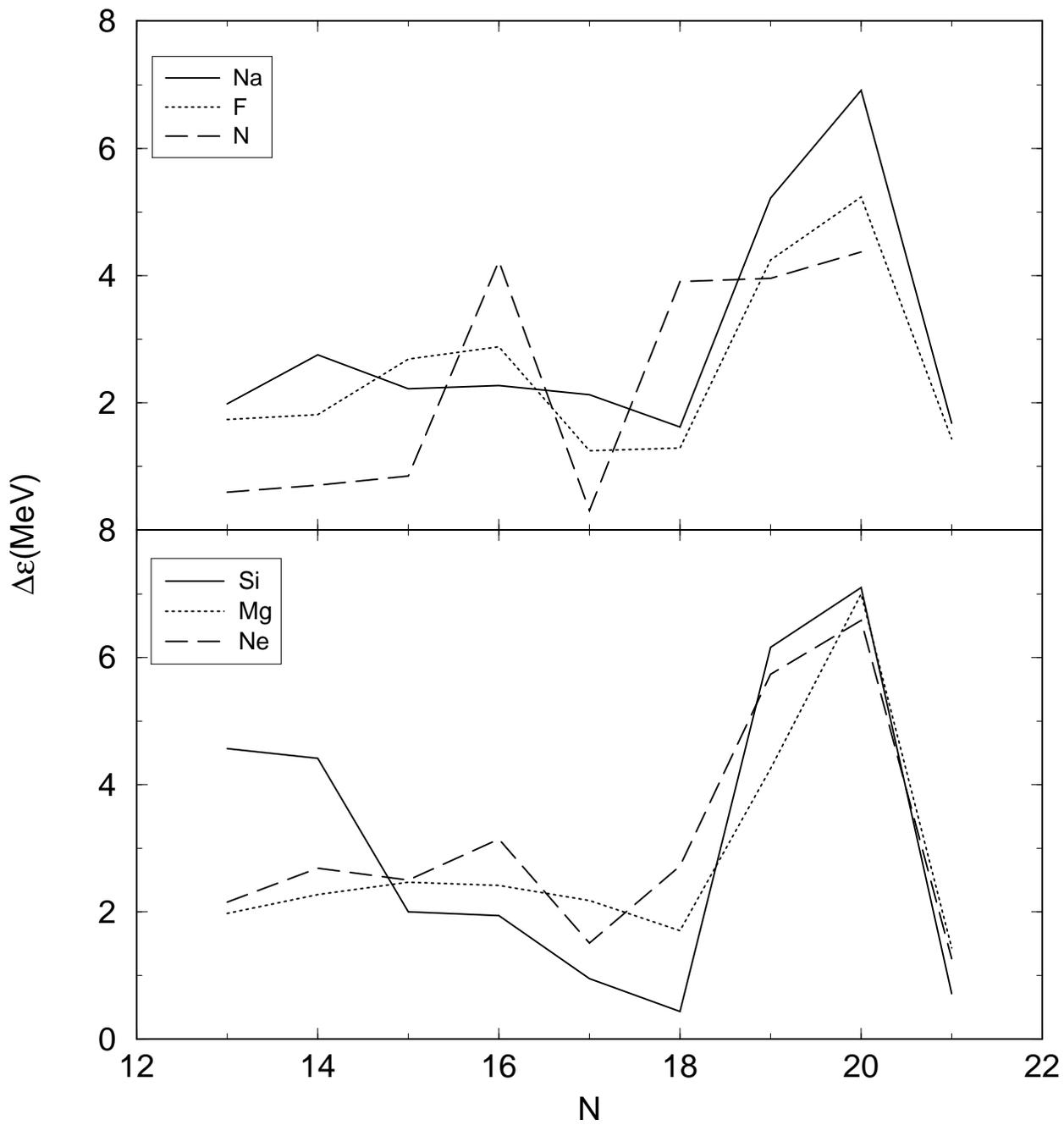
\item[Fig. 11] The same as in Fig. 10 for the energy gap $\Delta\epsilon$
across the Fermi surface. 
\end{itemize}

\end{document}